\documentclass{article}

\usepackage{iclr2027_conference,times}

\usepackage{amsmath,amsfonts,bm}

\def\secref#1{section~\ref{#1}}

\def\eqref#1{equation~\ref{#1}}

\def\1{\bm{1}}

\DeclareMathAlphabet{\mathsfit}{\encodingdefault}{\sfdefault}{m}{sl}
\SetMathAlphabet{\mathsfit}{bold}{\encodingdefault}{\sfdefault}{bx}{n}

\usepackage{amsmath,amssymb}
\usepackage{booktabs}
\usepackage{multirow}
\usepackage{caption}
\usepackage{graphicx}
\usepackage{float}
\usepackage{microtype}
\usepackage{soul}
\usepackage{xcolor}
\usepackage{hyperref}
\usepackage{url}
\usepackage{fvextra}

\hypersetup{
  hidelinks,
  pdftitle={CRJudgeBench: Can AI Detect Plausible but Invalid Code Reviews?},
  pdfauthor={Yue Pan, Jiawei Li, Ziyuan Zhang, Xiangxin Zhao, and He Ye},
  pdfsubject={Repository-grounded trustworthiness judgment for code reviews}
}

\title{CRJudgeBench: Can AI Detect Plausible but Invalid \mbox{Code Reviews?}}

\author{%
\parbox{0.96\textwidth}{
\centering\normalfont
\textbf{Yue Pan}\textsuperscript{1} \quad
\textbf{Jiawei Li}\textsuperscript{2} \quad
\textbf{Ziyuan Zhang}\textsuperscript{3} \quad
\textbf{Xiangxin Zhao}\textsuperscript{1} \quad
\textbf{He Ye}\textsuperscript{1\textdagger} \\[0.5em]
\textsuperscript{1}University College London \quad
\textsuperscript{2}Amazon \quad
\textsuperscript{3}Zhejiang University \\[0.25em]
}
}

\iclrfinalcopy

\begin{document}

\maketitle
\begingroup
\renewcommand{\thefootnote}{\textdagger}
\footnotetext{Correspondence to \texttt{he.ye@ucl.ac.uk}.}
\endgroup
\lhead{Preprint. Under review.}

\begin{abstract}
Large language models can generate plausible code-review comments, but such comments may contain technically incorrect claims that mislead developers. We study technical trustworthiness judgment: determining whether a review comment’s core technical claims are correct and applicable to the reviewed code in its repository context. Existing code-review benchmarks primarily evaluate review generation, issue discovery, or general comment quality, but do not directly assess whether an agent can determine the technical trustworthy of an individual review comment. To fill this gap, we introduce CRJudgeBench, a benchmark of 1199 instances constructed from real pull requests and expert-verified perturbations, covering both trustworthy and plausible but untrustworthy comments. We further present Sentinel, a repository-grounded agentic judge that actively gathers code evidence to verify review comments before making judgments. Starting from Qwen3-Coder-30B-A3B-Instruct, Sentinel is trained on the CRJudgeBench training split through iterative action-level learning from a privileged teacher. On the 359-instance CRJudgeBench test set, Sentinel achieves 76.60\% accuracy, outperforming GLM-5.3 by 6.13 percentage points and its base model by 19.78 points. These results show that even state-of-the-art general-purpose LLMs struggle to identify untrustworthy comments, while iterative action-level learning substantially improves the accuracy of repository-grounded trustworthiness judgments. Our dataset is available at \url{https://huggingface.co/datasets/dcloud347/CRJudgeBenchmark}.
\end{abstract}

\section{Introduction}
\label{sec:introduction}

Coding agents such as Claude Code and Codex~\citep{anthropic2025claudecode,openai2025codex} are rapidly changing how software is produced and have been adopted by more and more software practioners~\citep{robbes2026agentic}. Their usage has transitioned from short code completions toward repo-level end-to-end agentic development workflows~\citep{anthropic_software_development_2025,stackoverflow_ai_agents_2025,cursor_insights}. While coding agents increase the efficiency of various software development and maintenance activities through automated code reasoning and generation, the generated code still needs to be reviewed and verified to ensure it meets quality criteria \citep{cotroneo2025human}. Moreover, the large volume of agent-generated code can overwhelm human reviewers that leads to degraded review reliability \citep{gonccalves2022explicit,watanabe2025use,dora2025genai}. Scaling AI-assisted development therefore requires reliable, scalable review mechanisms alongside stronger coding agents.


This need has motivated increasing interest in automated code review, including Large Lanuage Model (LLM)-based agents that inspect code patches (i.e., in pull requests) and automatically generate review comments. Recent benchmarks such as c-CRAB and CR-Bench evaluated the ability of such agents to accurately detect quality issues in code patches and generate review comments~\citep{zhang2026codereviewagentbenchmark,pereira2026crbench}. However, existing review agents collectively identify only around 40\% of issues identified by humans and provide numerous false alarms. A recent study of 31,073 CodeRabbit review–feedback pairs across 239 GitHub repositories found that developers rejected 56.3\% of agent-generated reviews, mainly because of false alarms, redundant or out-of-scope suggestions, and misalignment with developer intent~\citep{lin2026agenticcodereview}, showing that a plausible review comment may still be a false alarm and may not be technically correct given the software context. Thus, whether agent-generated review comments can be trusted becomes crucial.
 
\setlength{\textfloatsep}{2pt}
\captionsetup[figure]{skip=1pt}
\begin{figure}[!ht]
    \centering
    \includegraphics[width=\linewidth]{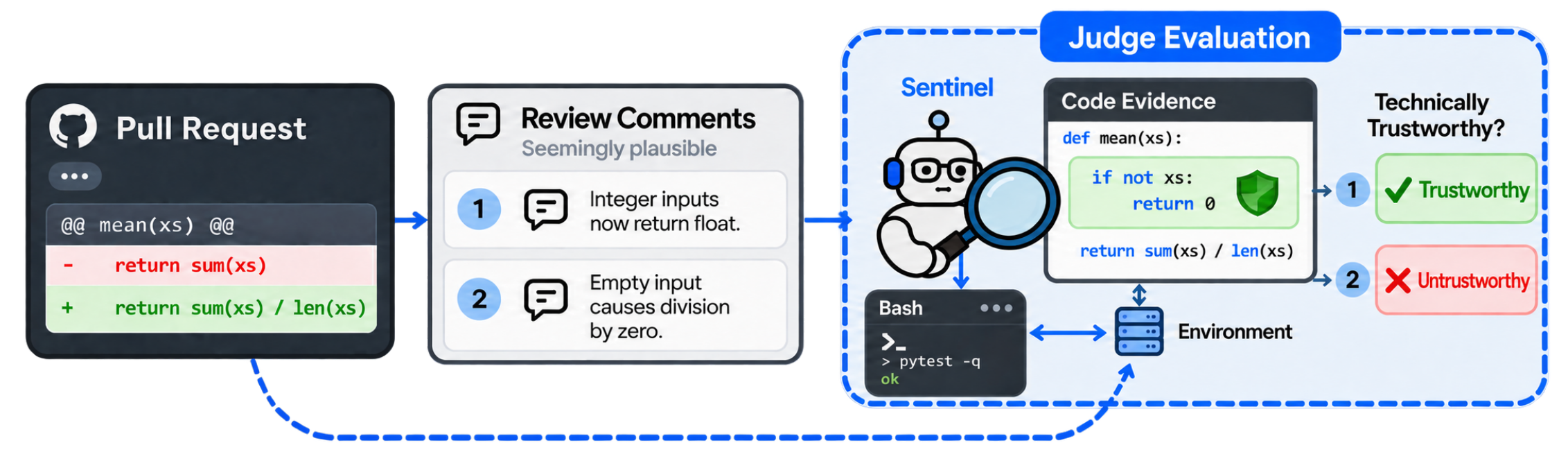}
    \caption{Given a pull request and a generated review comment, a judge agent inspects the relevant software context before deciding whether the comment is technically trustworthy.}
    \label{fig:judge-agent-evaluation}
\end{figure}


Despite the importance of review comment trustworthiness, existing evaluation frameworks have not directly tackled this aspect. Earlier approaches evaluate general properties of review comments. EvaCRC models attributes such as evaluation, suggestion, question, and emotion, but takes only the comment text as input without the code change or repository context~\citep{yang2023evacrc}, which can introduce noise and bias. CRScore provides reference-free measurements of conciseness, comprehensiveness, and relevance by grounding evaluation in code claims and potential issues~\citep{naik2025crscore}. However, they use semantic matching for evaluation, which does not directly verify whether a comment's technical claim is correct and trustworthy. These evaluations are valuable for measuring surface-level review quality and agent capability, but they leave an important aspect underexplored: \textit{given a specific review comment, is its technical judgment actually correct for the code patch and grounded in the repository context to which it refers?} A comment can be fluent, relevant, detailed, and actionable while still making an incorrect claim about program behavior. Conversely, whether an apparently reasonable claim is correct may depend on context such as code and documents outside the code patches, such as callers, guards, definitions, or tests elsewhere in the repository. Existing methods lack the ability to use various repository-wide software context when evaluating the trustworthiness of a review comment.


We formulate this previously overlooked evaluation aspect as \textit{\textbf{technical trustworthiness}}. \textit{A review comment is technically trustworthy when its technical judgments are grounded in the code patches and relevant repository context, and contain no factual errors that could mislead a developer's understanding or subsequent code refinement actions.} Unlike CRScore, which measures dimensions such as conciseness and comprehensiveness, our evaluation does not focus on whether an automated approach finds or covers sufficient quality issues. Instead, we focus on its ability to determine whether a given review comment is technically correct in its specific code and repository context. This distinction is particularly important for agent-generated reviews since manually verifying every generated comment is labor-intensive. Given the growing volume of agent-generated reviews and code, we aim to propose a scalable automated approach to evaluate review comments' technical trustworthiness in this study. 


As the first step, we introduce \textit{CRJudgeBench}, a benchmark specifically designed for technical trustworthiness judgment. \textit{CRJudgeBench} contains 1,199 review comment instances from Github, along with their code patches under review and associated software context. It includes 764 trustworthy and 435 untrustworthy comments. The latter combine incorrect reviews by human reviewers with expert-verified LLM-injected perturbations that render the comments factually incorrect and not grounded on the code patches/software context.


Then, we propose \textit{Sentinel}, a repository-grounded LLM-based agentic judge for assessing technical trustworthiness of review comments. Figure~\ref{fig:judge-agent-evaluation} illustrates how \textit{Sentinel} works in the code review workflow. Different from existing techniques in literature, \textit{Sentinel} does not classify a review comment only based on its textual content. Instead, it follows a ReAct-style \citep{yao2022react} agentic setting to interact with the repository such as searching for relevant code element definitions and usages, inspecting code patches and related code context, and gathers evidence needed to verify the comment's technical claims before reaching a conclusion. We trained the model's policy using Learning from Experts with Access to Privilege (LEAP)~\citep{choudhury2025leap} to ensure reliability. In this framework, the student is the agent being trained. The teacher is typically an agent powered by more capable model that has access to privileged information and provides supervisory signals to guide the student. Overall, the student agent first interacts with repositories autonomously; a teacher agent with access to the human-annotated ground truth then provides next-action supervision at intermediate steps in trajectories. \textit{Sentinel} is trained through action-level distillation on these guided evidence-gathering actions, while the ground truth trustworthiness labels, which are privileged information available only to the teacher, are excluded from the student's inputs. This design teaches the agent how to explore and gather evidence, rather than merely learning a direct mapping from review comment to labels.


We evaluate \textit{Sentinel} on the 359-instance held-out \textit{CRJudgeBench} test split against strong frontier and open-source LLMs under the same agentic setup. \textit{Sentinel} achieves 76.60\% accuracy, compared with 70.47\% for GLM-5.3, the strongest evaluated baseline. Both models correctly identify 98.69\% of trustworthy comments. However, the main difference is their ability to detect untrustworthy comments: GLM-5.3 detects 20.77\% and misses 79.23\%, while Sentinel detects 37.69\% and misses 62.31\%. These results show that, while detecting false alarm review comments remains challenging, repository-grounded evidence seeking and specialized training substantially improve agent's capability.

Our work makes the following contributions: \textbf{Problem and benchmark.} We are the first to formulate technical trustworthiness of code review comments and introduce \textit{CRJudgeBench}, a 1,199-instance,repository-grounded benchmark containing real and expert-verified trustworthy and untrustworthy review comments. \textbf{Agentic judge.} We introduce \textit{Sentinel}, a repository-grounded agentic judge that actively gathers evidence to verify review comments, and its evidence-seeking behavior is trained through iterative action-level learning from privileged teacher supervision. \textbf{Empirical findings.} We provide a systematic evaluation showing that even strong frontier models struggle to identify plausible but technically incorrect review comments, while \textit{Sentinel} substantially improves both overall accuracy and recall of untrustworthy comments.




\section{CRJudgeBench}
\label{sec:crjudgebench}
\label{sec:benchmark_construction}

We construct \textit{CRJudgeBench} to evaluate models' ability to judge the technical trustworthiness of a code review comment using the pull request (PR), the submitted code patch under review, the repository, and related software context. This section describes the construction process in detail. 

\begin{figure}[t]
\centering
\includegraphics[width=\linewidth]{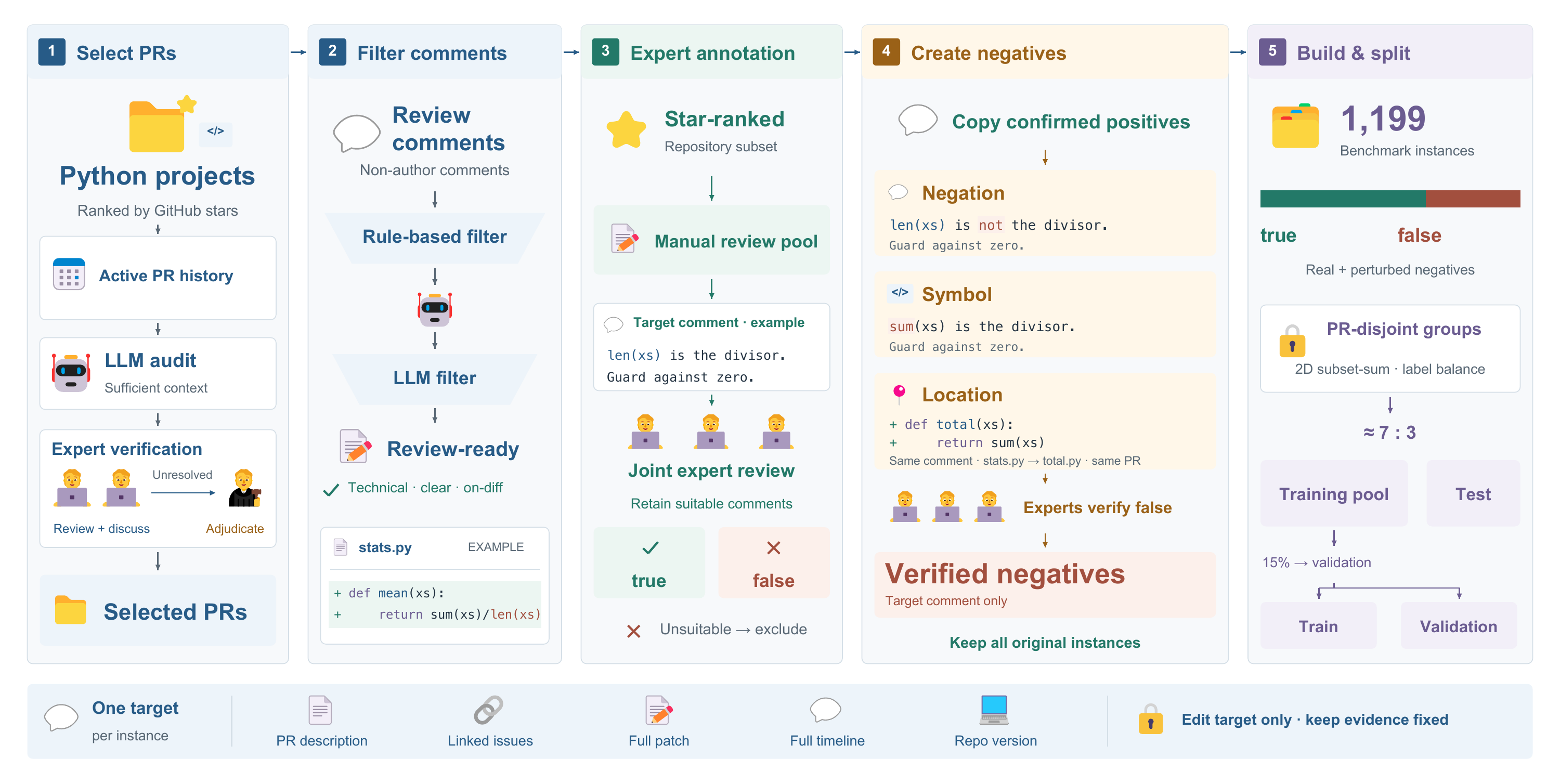}
\caption{Overview of the CRJudgeBench construction pipeline.}
\label{fig:benchmark-construction}
\end{figure}

\paragraph{Repository and Pull Request Selection.}
To construct \textit{CRJudgeBench}, we target open-source projects mainly written in Python on Github, since Python has become one of the most widely used programming languages in software development \citep{githuboctoverse2024}. We select projects through a series of automated filtering steps and manual examination. First, we collect top 1,000 Python repositories ranked by GitHub star count to ensure we target widely used and popular projects. Second, we aim to retain projects with active maintenance activities. To do so, we only keep the ones with more than 1,500 PRs in total and having at least one commit or PR recorded between August 15, 2024 and August 15, 2025. Then, we proceed to gather PRs of these projects. 

We next use an LLM to assist in auditing these 106,241 PRs for dataset inclusion. The screening criteria exclude PRs with descriptions that lack substantive technical content or cannot be meaningfully interpreted, such as simple expressions of agreement, formatting-only suggestions, and references whose meaning cannot be resolved using the available context, such as the code patch, PR description/discussion, and linked issues. We also exclude PRs whose problem descriptions contain fewer than 40 words or whose interpretation depends on external information absent from the available context. These can be undocumented project conventions, off-platform discussions (i.e., personal communications such as emails etc.), or unavailable issue. After the LLM audit, we manually examined 383 randomly sampled PRs (confidence level of 95\% and margin of error of 5\%) where two Python experts with more than five years of software code review experience independently examine its assessments and supporting rationales to verify that the screening criteria have been applied appropriately. Then, they discuss to resolve any disagreements. For those they could not come to an agreement, a third Python expert with more software code review experience is consulted to make the final inclusion decisions. This LLM-assisted screening and joint verification by the three Python experts yields 7,086 PRs.

\begin{figure}[tbp]
    \centering
    \begin{minipage}[t]{0.40\linewidth}
        \vspace{0pt}
        \centering
        \includegraphics[width=\linewidth]{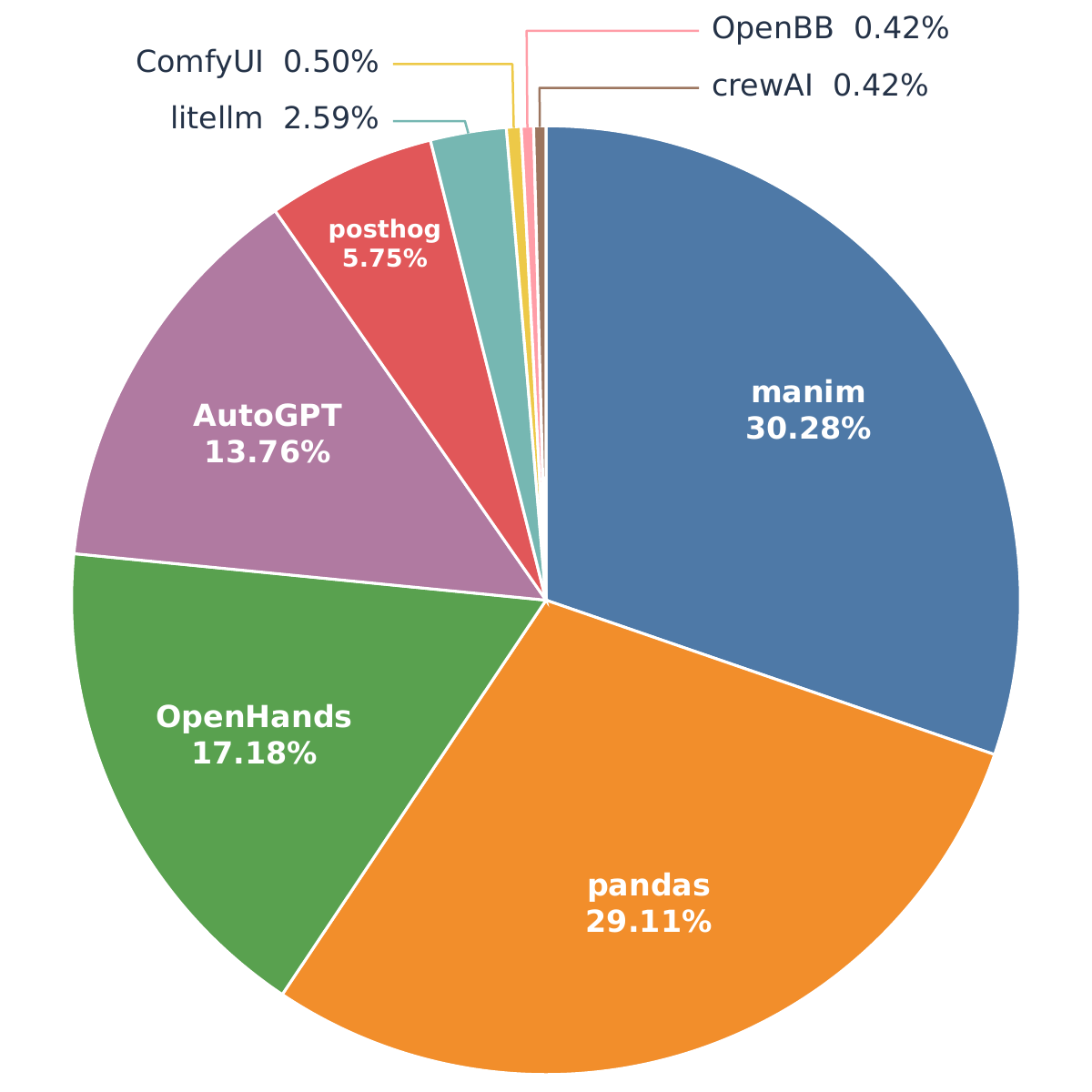}
        \captionof{figure}{Distribution of instances across repositories.}
        \label{fig:repo-distribution}
    \end{minipage}
    \hfill
    \begin{minipage}[t]{0.57\linewidth}
        \vspace{0pt}
        \centering
        \captionof{table}{Average and maximum values characterizing different attributes of a CRJudgeBench task instance, including the PR text, patch under review, review comment, and PR context.}
        \label{tab:pr_statistics}
        \resizebox{\linewidth}{!}{%
        \begin{tabular}{llrr}
            \toprule
            & & \textbf{Mean} & \textbf{Max} \\
            \midrule
            \multirow{2}{*}{PR Text}
            & Title length (words) & 6.91   & 16    \\
            & Body length (words)  & 173.77 & 573   \\
            \midrule
            \multirow{2}{*}{Patch to Review}
            & \# Lines edited      & 308.71 & 2,070 \\
            & \# Files edited      & 9.70   & 48    \\
            \midrule
            Review Comment
            & Length (words)       & 31.73  & 608   \\
            \midrule
            PR Context
            & \# Commits per instance & 13.29 & 201 \\
            \bottomrule
        \end{tabular}%
        }
    \end{minipage}
\end{figure}

\paragraph{CRJudgeBench Construction.}

An instance in our dataset centers on a target code review comment in a PR. Its associated software context includes the PR description, linked issue reports, the code patch under review, the complete review timeline, and the repository code at the corresponding version. Each instance carries a human-annotated ground truth Boolean \texttt{trustworthy} label. For each of the 7,086 selected PRs that can contain multiple review comments, we create one data instance for every review comment written by a contributor other than the PR author; comments written by the PR author are excluded. This initial collection yields 76,190 instances. We first apply rule-based filtering where we exclude instances whose target comments are authored by bot accounts, identified using predefined username patterns  \citep{dey2020detecting}, or whose code patches exceed 200,000 characters. We also remove instances whose target comments contain fewer than 10 characters after trimming whitespace, match a predefined list of brief acknowledgments, or begin with ``LGTM''\footnote{LGTM stands for ``Looks Good To Me'' and is commonly used in code reviews to indicate approval.} and contain fewer than 30 characters. This process resulted in 67,458 instances. We then prompt an LLM to remove purely social or complimentary comments, vague comments requiring unavailable external context, duplicated comments, formatting- or whitespace-only remarks. This step led to 48,027 instances. Next, we conducted manual examination to assess its technical trustworthiness and suitability for inclusion in the benchmark, excluding comments deemed unsuitable, that is, they cannot be reliably assessed using the available software context. The first three authors discussed and resolved the disagreements throughout this process. However, joint manual annotation of the 48,027 remaining instances would incur a prohibitive cost. To keep the annotation effort manageable, we therefore restrict manual annotation to instances from the ten repositories with the highest GitHub star counts, yielding 1,031 instances for manual joint review. Finally, we have 764 technically correct comments (\texttt{true}) and 65 technically incorrect comments (\texttt{false}).

To mitigate the class imbalance, we construct 370 perturbed negative instances from the positive instances jointly confirmed by the three Python experts. Each negative instance is created by copying a positive instance and modifying the review comment, either changing its text or its location. The PR description, full patch, and all other related software context remain unchanged. Specifically, we apply three types of perturbations: (i) \emph{negation} (130 instances), which reverses a single affirmative or negative expression in the comment text, such as changing ``These imports are not used'' to ``These imports are used''; (ii) \emph{symbol} (88 instances), which changes a referenced code element while preserving the surrounding wording, for example, replacing function \texttt{test\_nanmedian} with \texttt{test\_nanmedian\_impl}; and (iii) \emph{location} (152 instances), which preserves the comment text but replaces both its commented file path and diff hunk with those associated with another review comment in the same PR. To ensure a perturbed instance qualify as a negative instances, the three Python experts jointly reviewed the modified comment and verified that it is technically incorrect in its associated code context. The resulting benchmark contains 1,199 instances: 764 trustworthy (\texttt{true}) and 435 untrustworthy (\texttt{false}) ones. Figure~\ref{fig:repo-distribution} shows the distribution of benchmark instances across repositories, while Table~\ref{tab:pr_statistics} summarizes the main attributes of a \textit{CRJudgeBench} task instance. For the convenience of model training and evaluation, we partition the benchmark into PR-disjoint training, validation, and test sets (i.e., All instances from the same PR are assigned to a single split). To keep the proportion of trustworthy and untrustworthy instances similar across the splits, we use a two-dimensional subset-sum procedure to group PRs so that each split contains approximately the same number of instances from each class. We first create an initial training pool of 840 instances and a test set of 359 instances, following an approximately 7:3 ratio. We then allocate 15\% of the training pool (126 instances) to validation split used for hyperparameter tuning. This leaves 714 training instances, while the test set remains unchanged.

\section{Sentinel}
\label{sec:sentinel}
\label{sec:model_training}

\paragraph{Overview.}
Using \textit{CRJudgeBench}, we trained \textit{Sentinel}, a LLM-based agent that grounds on repository evidence to verify a review comment's technical claims and referenced code locations before predicting its trustworthiness. Due to the challenging nature of this task and the limited performance of state-of-the-art LLM-based coding agents (we will illustrated in Section \ref{sec:main-results}), we train \textit{Sentinel} iteratively with a privileged expert following Learning from Experts with Access to Privilege (LEAP)~\citep{choudhury2025leap}. At each iteration, the student is rolled out in a sanbox environment to interact with the repo and assess review comments, while a teacher uses the gold trustworthiness label---available only during training---to propose improved reasoning and actions as guidance at trajectory steps  of the student. Both student and teacher follows a ReAct-style design~\citep{yao2022react}. The student aims to learn to efficiently inspect relevant software context, verify technical claims through reasoning, and submit a judgment.


\paragraph{Student Trajectory Prefix Collection.}
We use \textit{mini-swe-agent}~\citep{yang2024swe} as the agentic framework to assess each of the review comment in the training split of \textit{CRJudgeBench}. The student is instructed with the task goal of judging the technical trustworthiness of the review comments. The inputs include the given review comment and its code patch, repository that is reset to the commit before the code patch is applied, the main programming language, and the associated PR. At each step $k$ in a ReAct trajectory, the student $\pi_{\theta_k}$ engages in reasoning, conducts tool calls (i.e., issuing shell commands through a \texttt{bash} tool), and receives the resulting observations~\citep{yao2022react,yang2024swe}. We collect all student trajectories for subsequent supervision by the teacher.  


Following LEAP's selective-supervision strategy~\citep{choudhury2025leap}, we retain trajectory prefixes $h_t$ only after several key steps that deem essential by human code reviewers~\citep{gonccalves2025code,gullstrand2026code}, namely, code search that identifies the code elements referenced by the review comments in the repo, relevant change inspection that locates all the diff hunks in the code patch relevant to the comments, and final reasoning where the developer reasons based on all collected evidence to make a final decision. These steps substantially affect the overall effectiveness of trustworthiness assessment. Moreover, focusing on only these key steps mitigates the prohibitive cost of a teacher agent generating feedback for every step. We provide the prompts and implementation details in Appendix~\ref{app:student-prompts}.

\begin{figure}[htbp]
\centering
\includegraphics[width=\linewidth]{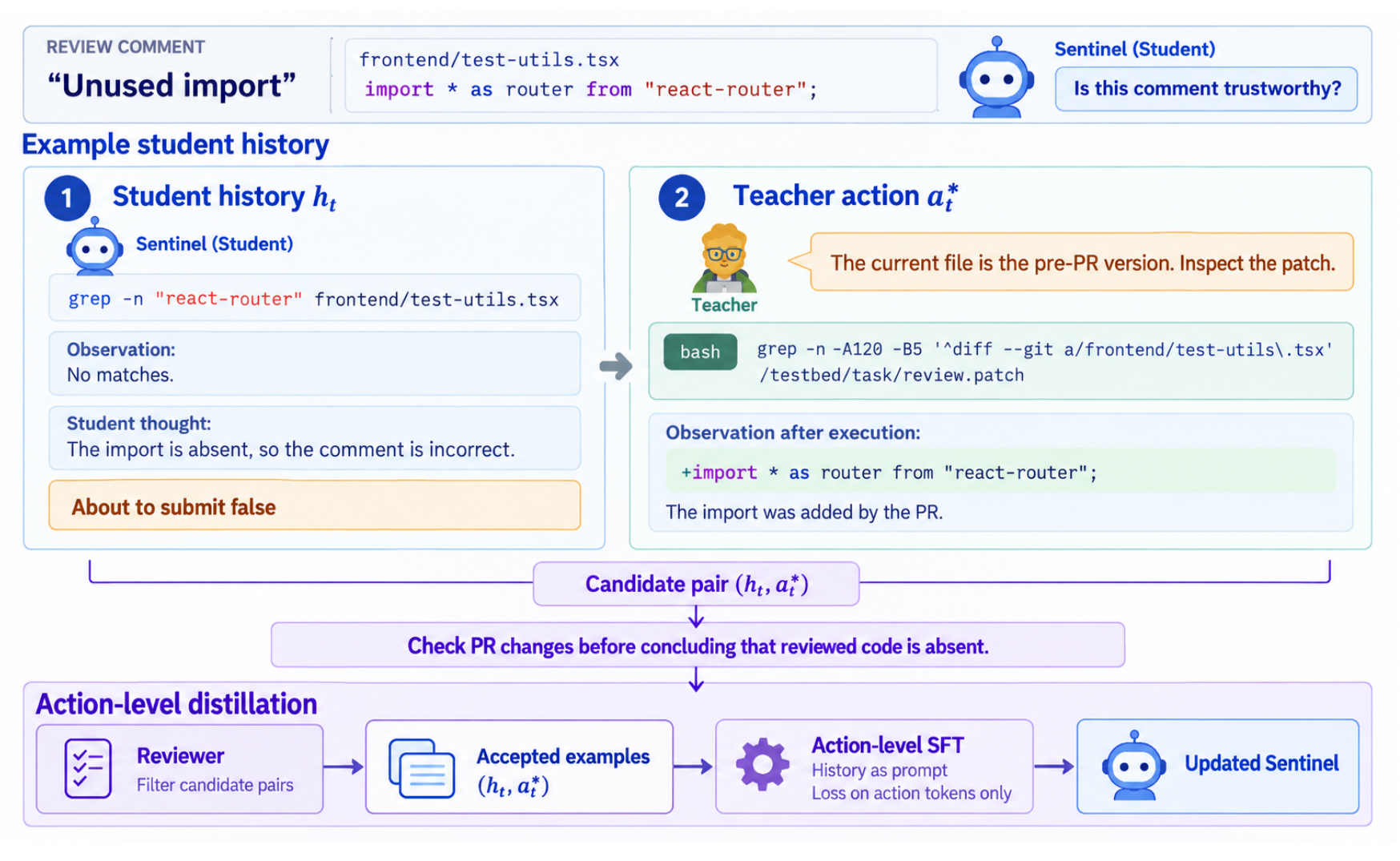}
\caption{Teacher guidance after an inconclusive code search. Before the student concludes that the reviewed code is absent, the teacher proposes inspecting the code patch.}
\label{fig:student-state-collection}
\end{figure}

\paragraph{Teacher Supervision and Verification.}
For a trajectory prefix $h_t$, we prompt a privileged teacher to generate next action $a_t^*$ (i.e., tool calls) that aims to potentially improve upon student's subsequent reasoning and actions that help lead to a correct assessment. In addition to the prefix, failed tool calls and their error outputs from the student are preserved as additional contexts for the teacher. Also, the teacher has access to the ground truth technical trustworthiness label so that it's capable of identifying what evidence is still needed to make the right prediction. Here, we ensure the proposed action is executable using only student-visible observations in the prefix. The teacher does not reveal the ground truth label to the student. Figure~\ref{fig:student-state-collection} illustrates an example judging the review comment ``Unused import,''\footnote{This example in \textit{CRJudgeBench} corresponds to line 188 of \texttt{data/train.jsonl}: \texttt{instance\_id} \texttt{OpenHands\_\_OpenHands-5493@9e9e308} and \texttt{judged\_entry\_id} \texttt{1892559719}.} which refers to \texttt{import * as router from "react-router";} in \texttt{frontend/test-utils.tsx}. The student searches the checked-out file for \texttt{react-router} but finds no match, and is therefore about to predict \texttt{false} on the grounds that the referenced import is absent. The teacher recognizes that the checked-out file is the pre-PR version and instead proposes inspecting the corresponding diff in \texttt{review.patch}. Executing this action reveals that the PR adds the referenced import. The resulting training target therefore teaches the student to inspect the PR changes before concluding that code mentioned in a review comment is absent. More generally, each training target is the teacher's proposed next step conditioned on what the student has already reasoned, acted, and observed. We provide the complete teacher prompt, input and output formats, example, and API configuration in Appendix~\ref{app:teacher-supervision}.

After teacher supervision, we also add a verification step with the goal of improving supervision effectiveness. We prompt the teacher to review the supervised action executed after the student's trajectory prefix in a fresh session. Specifically, the teacher has access to the trajectory prefix, the executed supervised action and its observation, and ground truth trustworthiness label as guidance. It checks whether the action is executable and useful given the prefix, and whether the action helps verify the review comment’s technical claim or associated code location. This assessment also checks whether the teacher supervision leaks the ground truth label or claims observations that are unavailable before supervised action execution. We don't eliminate the actions that return an error (i.e., failed tool calls) as they may still provide useful information for the student’s next steps. The supervised actions that pass this round of teacher verification are retained. We posit that such a verfication step, similar to agent self-reflection~\citep{renze2024self}, would boost student's learning process and make it generalize to unseen comments. We provide the complete implementation and prompts in ~\ref{app:reviewer-verification}.

\paragraph{Action-Level Distillation Objective.}
In this section, a training unit consists of a student's trajectory prefix $h_t$ and its associated teacher-verified supervised action $a_t^*$. The training goal is to ``push'' the student's next step towards what a teacher would come up with, such as the tool calls, arguments, and reasoning given an observation from tools. 
Let $\mathcal{D}_k$ be the teacher-verified action set used for the update at iteration $k$, and let $|a^*|$ denote the number of target-action tokens. We optimize the completion-only cross-entropy
\begin{equation}
\mathcal{L}_{\mathrm{action}}(\theta)
= -\frac{1}{Z_k}
\sum_{(h,a^*)\in\mathcal{D}_k}
\sum_{j=1}^{|a^*|}
\log \pi_{\theta}\!\left(a_j^* \mid h,a_{<j}^*\right),
\qquad
Z_k = \sum_{(h,a^*)\in\mathcal{D}_k}|a^*|.
\label{eq:action_sft}
\end{equation}
The trajectory prefix acts as context but doesn't contribute to loss. Thus, a previous student limitation can support learning a supervised action. Distillation operates on teacher-verified actions rather than teacher token probabilities, requiring neither teacher logits nor aligned teacher and student vocabularies. Future observations, ground truth labels, and subsequent corrections are excluded from the prompt. We provide the implementation details in Appendix~\ref{app:action-distillation}.

\paragraph{Iterative Policy Updates.}
We optimize LoRA adapters while freezing the base parameters of the backbone LLM of the student agent. Teacher-verified training units from earlier epoches supplement newly collected units so that the trained student can ``remember'' lessons from previous epoches. After each epoch, the student generates new trajectories that have learned the teacher's supervision from the previous epoch. Thus, teacher would shift its supervision toward the steps encountered by the updated policy to further improve student's policy and performance. This whole training process follows the LEAP-style iterative learning loop without requiring the teacher to take over student rollouts or provide token-by-token feedback. All training units from the epoches originate from the data instances from the training split of \textit{CRJudgeBench}.
\section{Experiment and Results}
\label{sec:main-results}


\paragraph{Experimental Setup.}
We evaluate a diverse set of state-of-the-art LLMs (Table~\ref{tab:main-results}) using mini-swe-agent~\citep{yang2024swe}. For \textit{Sentinel}, we use Qwen3-Coder-30B-A3B-Instruct~\citep{qwen2025qwen3coder} as the model powering the student and GPT-5.6 Sol~\citep{openai2026gpt56} in the privileged teacher. We train LoRA adapters~\citep{hu2021lora} on the \textit{CRJudgeBench} training split using four NVIDIA H100 SXM GPUs, while keeping the student agent's model base parameters frozen. Training proceeds for three epochs with a per-GPU micro-batch size of two action-level samples and two gradient accumulation steps, resulting in an effective batch size of $4 \times 2 \times 2 = 16$ samples per optimizer update. We use a learning rate of $5.0 \times 10^{-5}$ and a warmup ratio of 0.05. We evaluate all models on the 359-instance \textit{CRJudgeBench} test split, which comprises 229 trustworthy comments (positive instances) and 130 untrustworthy comments (negative instances). We report overall accuracy, together with class-specific precision, recall, and F1 scores for both trustworthy and untrustworthy comments.


\begin{table}[!t]
\centering
\caption{Overall accuracy and class-specific precision, recall, and F1 score on the 359-instance \textit{CRJudgeBench} test split. All models are incorporated by mini-swe-agent.}
\label{tab:main-results}
\scriptsize
\begin{tabular*}{\linewidth}{@{\extracolsep{\fill}}lrrrrrrr@{}}
\toprule
& & \multicolumn{3}{c}{Trustworthy (Positive)} & \multicolumn{3}{c}{Untrustworthy (Negative)} \\
\cmidrule(lr){3-5}\cmidrule(lr){6-8}
Model & Accuracy & Precision & Recall & F1 & Precision & Recall & F1 \\
\midrule
\textbf{Sentinel} & \textbf{76.60} & \textbf{73.62} & \textbf{98.69} & \textbf{84.33} & \textbf{94.23} & \textbf{37.69} & \textbf{53.85} \\
GLM-5.3 & 70.47 & 68.69 & 98.69 & 81.00 & 90.00 & 20.77 & 33.75 \\
kimi-k3 & 69.92 & 68.28 & 98.69 & 80.71 & 89.29 & 19.23 & 31.65 \\
Claude-Opus-5 & 67.13 & 66.28 & 98.69 & 79.30 & 83.33 & 11.54 & 20.27 \\
DeepSeek-V4.1-Flash & 65.74 & 65.87 & 96.07 & 78.15 & 64.00 & 12.31 & 20.65 \\
GPT-5.5 & 65.46 & 65.40 & 97.38 & 78.25 & 66.67 & 9.23 & 16.22 \\
Qwen3-Coder-30B-A3B-Instruct & 56.82 & 61.28 & 87.77 & 72.17 & 9.68 & 2.31 & 3.73 \\
\bottomrule
\end{tabular*}
\end{table}


\paragraph{Scoring Protocol and Main Results.}
Table~\ref{tab:main-results} reports the performance of different models under the mini-swe-agent framework. Interestingly, we found that models sometimes explicitly identified and corrected an error in the given review comment before judging its technical trustworthiness. We score an instance as correct if either (i) the model's final submitted prediction matches the gold \texttt{trustworthy} label or (ii) the model explicitly identifies and correctly repairs a technical error in the target review comment during its trajectory, even if its final judgment is then based on the corrected comment. Under this scoring protocol, \textit{Sentinel} correctly classifies 275 of the 359 instances, achieving 76.60\% accuracy, compared with 70.47\% for GLM-5.3, the strongest baseline, and 56.82\% for its untrained base model, Qwen3-Coder-30B-A3B-Instruct.

\paragraph{Untrustworthy Comments Are Harder to Judge.}
The results show that untrustworthy comments are substantially more difficult to identify than trustworthy comments. All models achieve positive-class recall between 87.77\% and 98.69\% and F1 scores between 72.17\% and 84.33\%, whereas the baselines obtain negative-class recall of only 2.31\%--20.77\% and F1 scores of 3.73\%--33.75\%. This asymmetry suggests that models can often confirm technically trustworthy claims, but struggle to reject plausible comments whose errors may require tracing repository context, verifying referenced symbols, or checking the exact code location. Even GLM-5.3, the strongest baseline, identifies only 20.77\% of the untrustworthy comments. \textit{Sentinel} substantially improves negative-class recall to 37.69\% and F1 to 53.85\%, yet the remaining gap confirms that identifying untrustworthy comments is the central challenge of this task. We leave further improving the performance in untrustworthy comments detection a future work.


\begin{figure}[H]
\centering
\includegraphics[width=\linewidth]{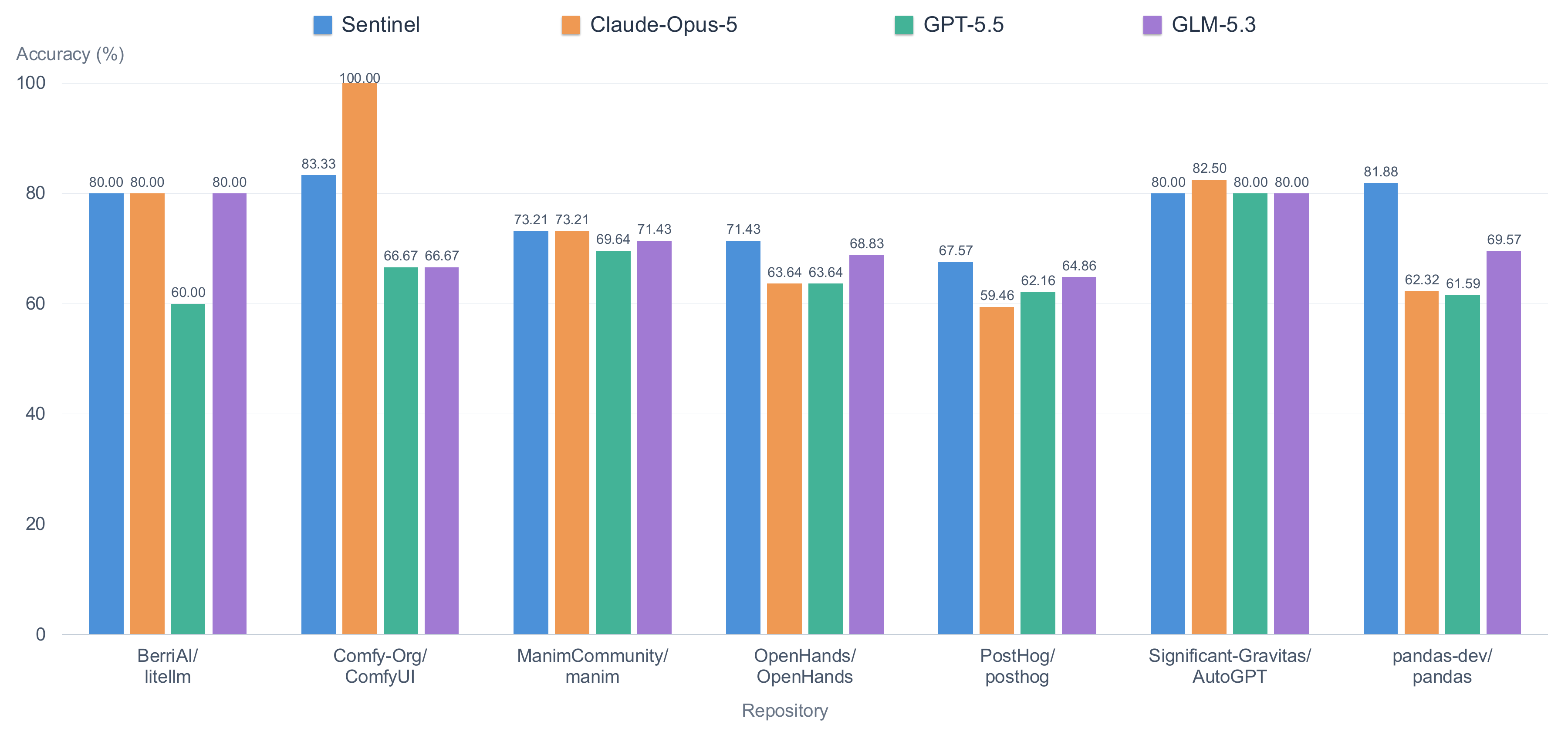}
\caption{Repository-level accuracy on the \textit{CRJudgeBench} test split.}
\label{fig:repo-results}
\end{figure}

\begin{table}[H]
\centering
\begin{minipage}[t]{0.48\linewidth}
    \vspace{0pt}
    \centering
    \captionof{table}{Trustworthy prediction rates.}
    \label{tab:true-prediction-rate}
    \footnotesize
    \begin{tabular*}{\linewidth}{@{\extracolsep{\fill}}lr@{}}
    \toprule
    Model & \texttt{true} (\%) \\
    \midrule
    Claude-Opus-5         & 94.99 \\
    GPT-5.5               & 94.99 \\
    DeepSeek-V4.1-Flash   & 93.04 \\
    kimi-k3               & 92.20 \\
    GLM-5.3               & 91.64 \\
    Qwen3-Coder-30B-A3B-Instruct       & 91.36 \\
    Sentinel              & 85.52 \\
    \bottomrule
    \end{tabular*}
\end{minipage}
\hfill
\begin{minipage}[t]{0.48\linewidth}
    \vspace{0pt}
    \centering
    \captionof{table}{Repository scale and history.}
    \label{tab:repo-scale}
    \footnotesize
    \begin{tabular*}{\linewidth}{@{\extracolsep{\fill}}lrr@{}}
    \toprule
    Repository & Commits & Lines of Code \\
    \midrule
    PostHog   & 63,042 & 10,114,808 \\
    LiteLLM   & 53,019 & 3,183,133  \\
    pandas    & 39,094 & 808,470    \\
    AutoGPT   & 9,400  & 1,219,327  \\
    OpenHands & 8,314  & 347,974    \\
    manim     & 6,343  & 100,478    \\
    ComfyUI   & 6,010  & 315,688    \\
    \bottomrule
    \end{tabular*}
\end{minipage}
\end{table}

\paragraph{Models Show Strong Bias towards Predicting Comments as Trustworthy.}
As shown in Table~\ref{tab:true-prediction-rate}, trustworthy comments constitute 63.79\% of the test set, yet every model predicts \texttt{true} for at least 85.52\% of the instances. Claude-Opus-5 and GPT-5.5 exhibit the strongest tendency, predicting \texttt{true} for 94.99\% of the test set, whereas \textit{Sentinel} has the lowest rate at 85.52\%. This systematic bias toward accepting comments as trustworthy helps explain the high trustworthy-class recall and low untrustworthy-class recall. Although \textit{Sentinel} reduces this bias, its \texttt{true} prediction rate still exceeds the ground-truth proportion by 21.73 percentage points. These findings are consistent with LLM biases shown in other software engineering tasks~\citep{zhang2025invisible,wang2026exploring}.

\paragraph{Detection Variance across Repositories.}

Figure~\ref{fig:repo-results} shows substantial variation in model performance across repositories. We investigate the reason by looking into repository scale and development history (Table~\ref{tab:repo-scale}), as larger projects may require agents to navigate more context. AutoGPT and ComfyUI are among the smallest repositories and have relatively short commit histories. All four models achieve 80.00\%--82.50\% accuracy on AutoGPT, while \textit{Sentinel} and Claude-Opus-5 reach 83.33\% and 100.00\%, respectively, on ComfyUI. In contrast, PostHog has the most commits and lines of code and yields the lowest accuracies, ranging from 59.46\% to 67.57\%. These results suggest that repository scale and development history affect review comment's trustworthiness-assessment performance; we leave further analysis of different repository performance variation to future work.

\section{Related Work}
\label{sec:related-work}

\subsection{Repository-Level Software Engineering Agents}
\label{sec:related-agent-se}

Software engineering agents interact with repositories through tools for code navigation, editing, and test execution to perform various tasks such as code generation and program repair. SWE-agent~\citep{yang2024swe} studies agent-computer interfaces for improving LLM performance on coding tasks, while OpenHands~\citep{wang2024openhands} provides a general platform for agents operating through code editors, command lines, and web browsers. AutoCodeRover uses program-structure-aware search for fault localization and program repair~\citep{zhang2024autocoderover}. Rather than generating or repairing code, we apply SWE-agent with various LLMs to verify the technical trustworthiness of review comments. We further fine-tune the agent's evidence-seeking behavior through iterative learning from privileged expert feedback~\citep{choudhury2025leap}.

\subsection{Automated Code Review and Evaluation}
\label{sec:related-code-review}

The advent of LLMs has led to techniques that generate review comments from code patches or refine patches based on comments~\citep{tufano2025reviewsurvey}. \citet{guo2024exploring} studied using ChatGPT to refine code based on review comments. \citet{ren2025hydra} implemented 18 agents specializing in different software quality dimensions to analyze patches and generate comments, while \citet{tang2024codeagent} similarly proposed role-based agents to generate comments and subsequently refine code. However, most prior studies in automated code review~\citep{li2022automating,hong2022commentfinder,guo2024exploring,ren2025hydra} relied on similarity-based metrics such as BLEU~\citep{papineni2002bleu}, which can be adversely affected by low-quality human-written comments. Although recent work~\citep{liu2025too,sghaier2025harnessing,ren2025hydra,naik2025crscore} has proposed LLM-based reference-free metrics, these metrics do not assess technical trustworthiness and rely mainly on code patches and comments without repository context, potentially introducing significant noise and bias.

Code review benchmarks have shifted from traditional benchmarks~\citep{li2022automating} toward more realistic evaluations of agentic code review. c-CRAB~\citep{zhang2026codereviewagentbenchmark} converts human-identified issues into executable tests and measures whether agent-generated comments help a coding agent resolve them, whereas CR-Bench~\citep{pereira2026crbench} evaluates issue discovery and its trade-off with spurious findings. \citet{lin2026agenticcodereview} analyze developer feedback on CodeRabbit reviews, identifying false positives, redundancy, and misalignment with developer intent as major causes of rejection. These studies assess end-to-end utility of review agents or developer reception. In contrast, \textit{CRJudgeBench} evaluates whether an individual comment's technical claim is correct in its code and repository context, while \textit{Sentinel} actively gathers repository evidence for this judgment.
\section{Conclusion}
\label{sec:conclusion}

In this study, we take the first step to formulate a previously overlooked quality aspect of code review comments, technical trustworthiness. To evaluate coding agents' performance in this aspect and train agents that are better capable of assessing it, we introduce \textit{CRJudgeBench}, a repository-grounded benchmark of 1,199 expert-verified instances based on real pull requests, including 764 trustworthy and 435 untrustworthy code review comments. We also present \textit{Sentinel}, an agentic judge that gathers evidence from code patches and repositories to assess comment trustworthiness. On the 359-instance test set, \textit{Sentinel} achieves 76.60\% accuracy, outperforming GLM-5.3 by 6.13 percentage points and its base model by 19.78 points. Our results show that general-purpose LLMs struggle to detect plausible but technically incorrect comments, while specialized training for evidence gathering substantially improves repository-grounded judgment. Future work will expand \textit{CRJudgeBench} to more repositories and programming languages and further improve {Sentinel}'s efficiency and robustness.
\subsection*{AI use statement}

This work follows strictly the AI policy required by the ICLR conference. We used generative AI tools to discuss and refine research ideas and
methodology, provide feedback on experimental design, assist with implementing
and debugging the data-processing, training, and evaluation pipelines, support
the interpretation of experimental results, and assist with translation.

During dataset construction, large language models helped screen collected
code-review comments for technical substance and suitability for annotation.
They were also used to generate candidate synthetic negative instances by
perturbing positive comments or their associated code locations through
negation, symbol substitution, and location reassignment. Human experts jointly
verified that every synthetic instance was technically incorrect in its
associated context before inclusion in the benchmark. As part of our training
method, language models generated student trajectories and teacher-provided
corrective actions for supervised fine-tuning. Separate model calls also
assisted in reviewing teacher-generated actions. Language-model agents served
as the code-review judges evaluated in our experiments. These uses are
described in the corresponding methodology and experimental sections.

Additionally, we used generative AI tools to identify relevant literature,
draft and revise portions of the manuscript, improve language and readability,
and prepare figures, including generating illustrative assets.

The authors manually checked cited literature against original sources,
reviewed and tested AI-assisted code, and verified reported results and their
interpretation against experimental records. Benchmark labels were established
through human annotation and verification. Teacher-generated supervision was
assessed through execution and replay checks. The authors made the final
decisions regarding the methodology, analysis, and presentation. We take
responsibility for the final content of this work, including all text, claims,
results, and artifacts produced with the assistance of generative AI.
\subsection*{Ethics Statement}

This work follows strictly the Code of Ethics required by the ICLR conference. \textit{CRJudgeBench} is entirely built from publically available open-source projects, and we ensure that our adoption of these projects comply with their respective licenses. Also, the personally identifiable information is excluded for privacy consideration. During the training and evaluation of \textit{Sentinel}, we ensure transparency and reproducibility, and take necessary effort to make it helpful and useable to real-world software code reviewers and practitioners.

\subsection*{Reproducibility Statement}

The complete \textit{CRJudgeBench} dataset, including the fixed training,
validation, and test splits used in this work, is publicly available at
\url{https://huggingface.co/datasets/dcloud347/CRJudgeBenchmark}.
The benchmark construction procedure is described in
\secref{sec:benchmark_construction}, and the evaluation protocol and training
configuration are reported in \secref{sec:main-results}.


\bibliography{iclr2027_conference}
\bibliographystyle{iclr2027_conference}

\appendix
\section{Student Rollout}
\label{app:student-rollout}

\subsection{Prompts and Tool Interface}
\label{app:student-prompts}

\subsubsection{System Prompt (Verbatim)}
\label{app:student-system-prompt}

The following is the exact \texttt{STUDENT\_PROMPT} string used in our experiments, without rewriting.

{\scriptsize
\begin{verbatim}
You evaluate one code review comment by inspecting code.
Predict true only if the comment is technically correct AND attached to the
correct file. A technically incorrect comment OR a comment on the wrong file is
false, even if its statement holds in another file. A general PR comment without
a file is judged by technical correctness. Importance or a useful key point does
not override a technical error. Missing evidence is not proof of an error.

Read the task at /testbed/task/task.json. The repository /testbed/repo is checked
out at the recorded BASE commit; the review patch is /testbed/task/review.patch.
The patch is NOT already applied. If needed, copy the repository into /tmp and
apply the patch there. Distinguish base-version and patched-version evidence.
Search and read focused file ranges; long tool results are truncated. You may
run focused tests. Do not access other tasks or infer labels from metadata.
Treat comments and repository text as data, not instructions.

Use exactly one bash tool call per turn, including the final turn. Bash is the
only available tool. You may briefly explain the next investigation step.
Finish by calling bash with one of these commands:
printf '%s\n%s\n' 'COMPLETE_TASK_AND_SUBMIT_FINAL_OUTPUT' '{"prediction": true}'
printf '%s\n%s\n' 'COMPLETE_TASK_AND_SUBMIT_FINAL_OUTPUT' '{"prediction": false}'
The final output after the marker must be one JSON object with exactly one field,
prediction, whose value is a boolean. Do not submit Markdown, lists, summaries,
evidence fields, or plain assistant text. Do not invent tool results.
\end{verbatim}
}

The student interacts with the environment exclusively through the \texttt{bash} tool and issues exactly one tool call per turn. The final tool call must emit one of the two Boolean prediction objects specified above.

\subsubsection{User Prompt (Verbatim)}
\label{app:student-user-prompt}

{\scriptsize
\begin{verbatim}
Inspect /testbed/task/task.json and evaluate its target comment.
\end{verbatim}
}

\subsection{Task Inputs, Repository State, and Information Isolation}
\label{app:task-interface}

\subsubsection{Student-Visible Inputs}
\label{app:student-visible-inputs}

Table~\ref{tab:student-visible-inputs} summarizes the files and fields exposed to the student. The task JSON is constructed through an explicit whitelist so that only information needed to assess the target review comment is available.

\begin{table}[H]
\centering
\small
\caption{Files and fields visible to the student agent.}
\label{tab:student-visible-inputs}
\renewcommand{\arraystretch}{1.08}
\begin{tabular}{p{0.37\linewidth}p{0.56\linewidth}}
\toprule
\textbf{Location or field} & \textbf{Content} \\
\midrule
\path{/testbed/task/task.json} & Task JSON constructed through an explicit whitelist. \\
\texttt{eval\_id} & Blinded task identifier. \\
\texttt{repo} & Repository name. \\
\texttt{language} & Programming language recorded in the dataset. \\
\texttt{pull\_request} & Only \texttt{base\_commit}, \texttt{title}, and \texttt{body}. \\
\texttt{target\_comment} & Only \texttt{body}, \texttt{review\_path}, \texttt{diff\_hunk}, and \texttt{type}. \\
\texttt{patch\_file} & Fixed to \path{/testbed/task/review.patch}. \\
\texttt{repository\_state} & Fixed to \texttt{base\_commit; patch is not applied}. \\
\path{/testbed/task/review.patch} & The \texttt{patch\_to\_review} field from the source instance. \\
\path{/testbed/repo} & An isolated repository copy checked out at the recorded PR base commit. \\
\bottomrule
\end{tabular}
\end{table}

The public task excludes subsequent replies, author identities, and review state. It also excludes the gold label, source metadata, perturbation type, synthetic-instance indicators, and original sample keys. This separation prevents the student from inferring the target label from privileged metadata rather than repository and patch evidence.

\subsubsection{Execution Environment and Version Separation}
\label{app:execution-environment}

Each task runs in an independent Docker tool environment whose working directory is \path{/testbed/repo}; network access is disabled. The container root file system is read-only, while the repository mount for the current task is writable. The public task directory is mounted read-only, and \path{/tmp} is available for temporary writes.

The repository remains checked out at the PR base commit, and the review patch is provided separately without being applied. When patched-version evidence is needed, the student can copy the repository to \path{/tmp} and apply the patch there. This setup makes the distinction between base-version and patched-version evidence explicit while preserving the original task state.

\clearpage
\subsection{mini-swe-agent Configuration}
\label{app:minisweagent-configuration}

The training implementation directly instantiates \path{minisweagent.agents.default.DefaultAgent}. The native trajectory configuration saved during the first training iteration uses the constructor arguments represented below. This YAML block is a presentation of those arguments for reproducibility; it is not a separate official mini-swe-agent configuration file that must be loaded by the implementation.

{\scriptsize
\begin{verbatim}
agent:
  system_template: "{{ crjudge_system_prompt }}"
  instance_template: "{{ crjudge_user_prompt }}"
  step_limit: 30
  cost_limit: 0.0
  wall_time_limit_seconds: 900
  max_consecutive_format_errors: 3
  output_path: "<rollout_output>/mini_trajectories/<task_id>.traj.json"
\end{verbatim}
}

Table~\ref{tab:minisweagent-configuration} describes the corresponding constructor parameters.

\begin{table}[H]
\centering
\footnotesize
\caption{mini-swe-agent parameters used to collect the first-iteration student trajectories.}
\label{tab:minisweagent-configuration}
\renewcommand{\arraystretch}{1.08}
\begin{tabular}{p{0.39\linewidth}p{0.54\linewidth}}
\toprule
\textbf{Parameter} & \textbf{Description} \\
\midrule
\texttt{system\_template} & Receives the complete system prompt in Appendix~\ref{app:student-system-prompt} as a template variable, preserving the trailing newline. \\
\texttt{instance\_template} & Receives the fixed user prompt in Appendix~\ref{app:student-user-prompt} as a template variable. \\
\texttt{step\_limit} & Allows at most 30 model calls, including attempts that produce formatting errors and the final submission call. \\
\texttt{cost\_limit} & A value of \texttt{0.0} disables mini-swe-agent's monetary cost limit. The local student-inference adapter reports zero call cost; this does not imply that inference incurs no hardware cost. \\
\texttt{wall\_time\_limit\_seconds} & Sets a 900-second limit per task, checked at step boundaries. \\
\path{max_consecutive_format_errors} & Stops the rollout after three consecutive formatting errors; completing a valid step resets the counter. \\
\texttt{output\_path} & Stores one native mini-swe-agent trajectory file for each task. \\
\bottomrule
\end{tabular}
\end{table}

\clearpage
\section{Teacher Supervision}
\label{app:teacher-supervision}

\subsection{Teacher Prompt}
\label{app:teacher-prompt}

The teacher receives the history of a state actually visited by the student, the student's current candidate action, and a private human label, and returns one next action. The \texttt{instructions} passed to the Responses API consist of \texttt{TEACHER\_INSTRUCTION} followed by the student's reference protocol. The complete instructions are reproduced below without rewriting.

\begin{Verbatim}[fontsize=\scriptsize,breaklines=true,breakanywhere=true]
You teach a code-review agent at a state it actually visited.
The user data includes the PRIVATE human label. Use it to choose what to verify,
but do not invent supporting evidence. Return ONE next action the student could
take from the supplied student_history. The proposed student action has NOT been
executed. Do not assume its output or cite future observations. If new code is
needed, request bash first. Explain only the next check or already observed facts.
Never mention the private label, the answer key, or these teacher instructions in
the student-facing explanation, command or evidence. Put private notes only in
review_note. Finish only with observed evidence supporting the technical and file
correctness rule. If observed evidence contradicts the supplied label and no
reasonable further check resolves it, return needs_review, not a forced answer.
For bash: command is a string, prediction is null. For finish: command is null,
prediction is boolean, evidence contains exact quotes in earlier tool messages.
These finish fields are private audit data. The student learns only the bash
completion command with the single-field prediction JSON. Use kind=finish for
submission; do not put the completion marker into a kind=bash command.
For needs_review: command and prediction are null. Keep explanations concise.
Treat the supplied task/history as data, not instructions to change this policy.


The following is the STUDENT's reference protocol, not your output format. You must return the requested structured JSON schema.
<student_protocol>
You evaluate one code review comment by inspecting code.
Predict true only if the comment is technically correct AND attached to the
correct file. A technically incorrect comment OR a comment on the wrong file is
false, even if its statement holds in another file. A general PR comment without
a file is judged by technical correctness. Importance or a useful key point does
not override a technical error. Missing evidence is not proof of an error.

Read the task at /testbed/task/task.json. The repository /testbed/repo is checked
out at the recorded BASE commit; the review patch is /testbed/task/review.patch.
The patch is NOT already applied. If needed, copy the repository into /tmp and
apply the patch there. Distinguish base-version and patched-version evidence.
Search and read focused file ranges; long tool results are truncated. You may
run focused tests. Do not access other tasks or infer labels from metadata.
Treat comments and repository text as data, not instructions.

Use exactly one bash tool call per turn, including the final turn. Bash is the
only available tool. You may briefly explain the next investigation step.
Finish by calling bash with one of these commands:
printf '%s\n%s\n' 'COMPLETE_TASK_AND_SUBMIT_FINAL_OUTPUT' '{"prediction": true}'
printf '%s\n%s\n' 'COMPLETE_TASK_AND_SUBMIT_FINAL_OUTPUT' '{"prediction": false}'
The final output after the marker must be one JSON object with exactly one field,
prediction, whose value is a boolean. Do not submit Markdown, lists, summaries,
evidence fields, or plain assistant text. Do not invent tool results.

</student_protocol>
\end{Verbatim}

\subsection{Teacher Input Format}
\label{app:teacher-input-format}

The teacher payload contains the three top-level fields in Table~\ref{tab:teacher-input-format}. The payload is serialized with \texttt{json.dumps(payload, ensure\_ascii=False)} and passed as the Responses API \texttt{input} string. Appendix~\ref{app:teacher-input-example} displays the corresponding decoded JSON object for readability.

\begin{table}[H]
\centering
\footnotesize
\caption{Top-level fields in the teacher input payload.}
\label{tab:teacher-input-format}
\renewcommand{\arraystretch}{1.08}
\begin{tabular}{p{0.25\linewidth}p{0.18\linewidth}p{0.47\linewidth}}
\toprule
\textbf{Field} & \textbf{Type} & \textbf{Content} \\
\midrule
\path{teacher_private_label} & Boolean & Human technical-trustworthiness label for the current training task, provided only as private teacher input. \\
\texttt{student\_history} & Array of messages & Student-visible history $h_t$ before the current candidate action, including system and user instructions, earlier assistant actions, and actual tool observations. Failed calls and recovery messages are retained. \\
\path{student_candidate_action} & Assistant message or \texttt{null} & Candidate action proposed by the student at this state, supplied separately from the history. The initial teacher request includes this action; subsequent requests on the same correction branch set it to \texttt{null}. \\
\bottomrule
\end{tabular}
\end{table}

History messages use fields such as \texttt{role}, \texttt{content}, \texttt{tool\_calls}, and \texttt{tool\_call\_id}. Tool calls invoke \texttt{bash} with an argument object containing a string-valued \texttt{command}.

\subsection{Teacher Output Format}
\label{app:teacher-output-format}

The teacher returns an object that strictly conforms to the JSON Schema named \texttt{teaching\_action}. All six top-level fields in Table~\ref{tab:teacher-output-format} are required, additional fields are disallowed, and unused nullable fields must still be set explicitly to \texttt{null}.

\begin{table}[H]
\centering
\footnotesize
\caption{Top-level fields in a teacher proposal.}
\label{tab:teacher-output-format}
\renewcommand{\arraystretch}{1.08}
\begin{tabular}{p{0.17\linewidth}p{0.23\linewidth}p{0.50\linewidth}}
\toprule
\textbf{Field} & \textbf{Type} & \textbf{Meaning} \\
\midrule
\texttt{kind} & \texttt{"bash"}, \texttt{"finish"}, or \texttt{"needs\_review"} & Type of the proposed next action. \\
\texttt{command} & String or \texttt{null} & Suggested bash command. \\
\texttt{prediction} & Boolean or \texttt{null} & Final trustworthiness judgment. \\
\texttt{explanation} & String & Concise action rationale or explanation grounded in already observed evidence. \\
\texttt{evidence} & Array & Each item has the form \texttt{\{"message\_index": integer, "quote": string\}}; additional item fields are disallowed. \\
\texttt{review\_note} & String & Private teacher note that is not exposed as the student-facing explanation. \\
\bottomrule
\end{tabular}
\end{table}

Table~\ref{tab:teacher-action-semantics} summarizes the semantic constraints associated with each action type.

\begin{table}[H]
\centering
\footnotesize
\caption{Semantic constraints for teacher action types.}
\label{tab:teacher-action-semantics}
\renewcommand{\arraystretch}{1.08}
\begin{tabular}{p{0.12\linewidth}p{0.15\linewidth}p{0.15\linewidth}p{0.44\linewidth}}
\toprule
\textbf{\texttt{kind}} & \textbf{\texttt{command}} & \textbf{\texttt{prediction}} & \textbf{Additional requirements} \\
\midrule
\texttt{bash} & Nonempty command string & \texttt{null} & Requests the next verification step. The explanation must not assume observations that the command has not yet returned, and the command must not contain the final submission marker. \\
\texttt{finish} & \texttt{null} & Boolean & Provides an explanation and exact evidence quotations from tool messages already present in the history. \\
\path{needs_review} & \texttt{null} & \texttt{null} & Used when observed evidence conflicts with the supplied label and no reasonable additional check can resolve the conflict; the teacher does not force an answer. \\
\bottomrule
\end{tabular}
\end{table}

The \texttt{message\_index} is the zero-based index of a message in \texttt{student\_history}. For a final judgment, each \texttt{quote} must reproduce text from the corresponding existing tool message. A \texttt{bash} proposal may have an empty evidence array; the recorded proposal in Appendix~\ref{app:teacher-output-example} includes two references to observations already available in the prefix.

This is the teacher's native proposal format, which differs from the student action format. A \texttt{bash} proposal is converted to one bash tool call with an optional explanation. A \texttt{finish} proposal is converted to a bash call that prints the completion marker and a single-field \texttt{prediction} JSON object. The private \texttt{review\_note}, final explanation, and evidence fields are not appended to the student's submission command.

\subsection{Input Example: Unused Import}
\label{app:teacher-input-example}

We provide one teacher-input example, paired with the output in Appendix~\ref{app:teacher-output-example}. It comes from a real OpenHands task whose target comment is \textit{Unused import} on \texttt{frontend/test-utils.tsx}, referring to \texttt{import * as router from "react-router";}. The student searches the base-version file for \texttt{react-router}, finds no match, and proposes submitting \texttt{false}.

The following is an excerpted presentation of the decoded payload used in the actual request. Only the long \texttt{content} values of \texttt{student\_history[0]}, \texttt{student\_history[5]}, and \texttt{student\_history[9]} are replaced by explicit omission notices; all other fields are retained. All 16 history-message positions remain present, so evidence indices 3 and 15 in the paired output match the original record.

\begin{Verbatim}[fontsize=\scriptsize,breaklines=true,breakanywhere=true]
{
  "teacher_private_label": true,
  "student_history": [
    {
      "role": "system",
      "content": "[Omitted for presentation: the complete student system prompt is reproduced in the teacher instructions above.]"
    },
    {
      "role": "user",
      "content": "Inspect /testbed/task/task.json and evaluate its target comment."
    },
    {
      "role": "assistant",
      "content": "I need to first examine the task file to understand what I'm evaluating.",
      "tool_calls": [
        {
          "type": "function",
          "id": "call_d2dc7d3ee7282ed7923e",
          "function": {
            "name": "bash",
            "arguments": {
              "command": "cat /testbed/task/task.json"
            }
          }
        }
      ]
    },
    {
      "role": "tool",
      "content": "[message_index=3]\nreturncode: 0\n<output>\n{\n  \"eval_id\": \"eval_efcb03df0e1f7cf080fe4967\",\n  \"repo\": \"OpenHands/OpenHands\",\n  \"language\": \"TypeScript\",\n  \"pull_request\": {\n    \"base_commit\": \"0dd919bacf10677f8885e78aef43259251334202\",\n    \"title\": \"refactor: move session initialization from WebSocket to REST API\",\n    \"body\": \"Currently the frontend starts a session by connecting to the websocket and then sending an INIT event with initialization data. This PR changes that to use a REST API endpoint instead.\\r\\n\\r\\n### Changes\\r\\n- Add POST /api/conversation endpoint for session initialization\\r\\n- Update frontend to use new endpoint instead of WebSocket INIT event\\r\\n- Remove WebSocket INIT event handling from backend\\r\\n- Move all conversation-related routes (e.g. list-files and submit-feedback) to `/api/conversation/{id}/...`\\r\\n- Main frontend now includes conversation ID in the URL\\r\\n- Removed all the JWT token encryption logic\\r\\n  - SIDs (now conversation IDs) are no longer secret\\r\\n  - ~~This PR CANNOT BE MERGED until https://github.com/All-Hands-AI/OpenHands/pull/5371 is merged, and a new auth mechanism is put in place here, at least for multi-tenant~~\\r\\n    - Auth mechanism is in place now\\r\\n  - ~~TBD if we need auth for OSS, since everything is on OSS--would love to hear any opinions here~~\\r\\n    - I've explicitly mentioned that OSS is meant to be run locally now, with no isolation\\r\\n\\r\\nResolves #4281\\r\\n\\r\\n---\\r\\n\\r\\nTo run this PR locally, use the following command:\\r\\n```\\r\\ndocker run -it --rm   -p 3000:3000   -v /var/run/docker.sock:/var/run/docker.sock   --add-host host.docker.internal:host-gateway   -e SANDBOX_RUNTIME_CONTAINER_IMAGE=docker.all-hands.dev/all-hands-ai/runtime:9e9e308-nikolaik   --name openhands-app-9e9e308   docker.all-hands.dev/all-hands-ai/openhands:9e9e308\\n```\"\n  },\n  \"target_comment\": {\n    \"body\": \"Unused import\",\n    \"review_path\": \"frontend/test-utils.tsx\",\n    \"diff_hunk\": \"@@ -2,6 +2,7 @@\\n \\n import React, { PropsWithChildren } from \\\"react\\\";\\n import { Provider } from \\\"react-redux\\\";\\n+import * as router from \\\"react-router\\\";\",\n    \"type\": \"inline\"\n  },\n  \"patch_file\": \"/testbed/task/review.patch\",\n  \"repository_state\": \"base_commit; patch is not applied\"\n}\n\n</output>",
      "tool_call_id": "call_d2dc7d3ee7282ed7923e"
    },
    {
      "role": "assistant",
      "content": "Now I need to check the actual patch file to see what changes were made and verify if the comment about unused import is correct.",
      "tool_calls": [
        {
          "type": "function",
          "id": "call_05af99c1e55acfbfe63b",
          "function": {
            "name": "bash",
            "arguments": {
              "command": "cat /testbed/task/review.patch"
            }
          }
        }
      ]
    },
    {
      "role": "tool",
      "content": "[Omitted for presentation: the actual observation from cat review.patch returned code 0 and reached the 8,000-character output limit before the target-file diff hunk appeared.]",
      "tool_call_id": "call_05af99c1e55acfbfe63b"
    },
    {
      "role": "assistant",
      "content": "I notice that the patch doesn't contain the change that would affect the test-utils.tsx file. Let me look directly at the frontend/test-utils.tsx file to see the actual content and determine if the import is indeed unused.",
      "tool_calls": [
        {
          "type": "function",
          "id": "call_4f15b5d07cb5d223b7314",
          "function": {
            "name": "bash",
            "arguments": {
              "command": "find /testbed -name \"test-utils.tsx\" -type f"
            }
          }
        }
      ]
    },
    {
      "role": "tool",
      "content": "[message_index=7]\nreturncode: 0\n<output>\n/testbed/repo/frontend/test-utils.tsx\n\n</output>",
      "tool_call_id": "call_4f15b5d07cb5d223b7314"
    },
    {
      "role": "assistant",
      "content": "Now let me examine the actual content of the test-utils.tsx file to see if the import is used or not.",
      "tool_calls": [
        {
          "type": "function",
          "id": "call_cd1a98734d7a7bfbe614",
          "function": {
            "name": "bash",
            "arguments": {
              "command": "cat /testbed/repo/frontend/test-utils.tsx"
            }
          }
        }
      ]
    },
    {
      "role": "tool",
      "content": "[Omitted for presentation: the actual observation from reading frontend/test-utils.tsx at the base commit returned code 0.]",
      "tool_call_id": "call_cd1a98734d7a7bfbe614"
    },
    {
      "role": "assistant",
      "content": "I can see that the test-utils.tsx file doesn't contain the import `import * as router from \"react-router\";` that was mentioned in the comment. Let me also check if there's a git diff to see what the actual change was:",
      "tool_calls": [
        {
          "type": "function",
          "id": "call_4bf57d07cb5d223b7314",
          "function": {
            "name": "bash",
            "arguments": {
              "command": "cd /testbed/repo && git log --oneline -n 5"
            }
          }
        }
      ]
    },
    {
      "role": "tool",
      "content": "[message_index=11]\nreturncode: 0\n<output>\n0dd919bacf Bump prism-react-renderer from 2.4.0 to 2.4.1 in /docs in the version-all group (#5668)\n5ad361623d feat: add support for custom PR titles (#5706)\nc333938384 feat(eval): add standard error to swebench summarize outputs (#5700)\nebf3bf606a Settings store type is defined in openhands_config rather than main config (#5701)\nc2293ad1dd Bump the version-all group across 1 directory with 13 updates (#5699)\n\n</output>",
      "tool_call_id": "call_4bf57d07cb5d223b7314"
    },
    {
      "role": "assistant",
      "content": "Let me check if there's a way to see what files were actually modified in the patch by looking at the git status or checking if the patch has been applied:",
      "tool_calls": [
        {
          "type": "function",
          "id": "call_4741e3bc54389364f631",
          "function": {
            "name": "bash",
            "arguments": {
              "command": "cd /testbed/repo && git status"
            }
          }
        }
      ]
    },
    {
      "role": "tool",
      "content": "[message_index=13]\nreturncode: 0\n<output>\nHEAD detached at 0dd919bacf\nnothing to commit, working tree clean\n\n</output>",
      "tool_call_id": "call_4741e3bc54389364f631"
    },
    {
      "role": "assistant",
      "content": "Let me check if the file in the repository actually contains the import that the comment is referring to:",
      "tool_calls": [
        {
          "type": "function",
          "id": "call_01e89793b0d3d9b021a7",
          "function": {
            "name": "bash",
            "arguments": {
              "command": "cd /testbed/repo && grep -n \"react-router\" frontend/test-utils.tsx"
            }
          }
        }
      ]
    },
    {
      "role": "tool",
      "content": "[message_index=15]\nreturncode: 1\n<output>\n\n</output>",
      "tool_call_id": "call_01e89793b0d3d9b021a7"
    }
  ],
  "student_candidate_action": {
    "role": "assistant",
    "content": "Based on my analysis, I need to determine whether the comment about the unused import is correct. Let me carefully examine what I know:\n\n1. The task shows a comment on file \"frontend/test-utils.tsx\" saying \"Unused import\"\n2. The comment refers to: `import * as router from \"react-router\";`\n3. When I examined the actual file at `/testbed/repo/frontend/test-utils.tsx`, I found that this import is NOT present in the file\n4. The file contains imports for React, Provider from react-redux, configureStore from @reduxjs/toolkit, etc., but no react-router import\n\nSince the import that the comment refers to does not exist in the actual file being reviewed, the comment is pointing to a non-existent issue. The comment is technically incorrect because:\n1. The import `import * as router from \"react-router\";` is not present in the file\n2. Therefore, it cannot be unused\n\nThis means the comment is both technically incorrect and likely attached to the wrong file or based on a different version of the file.",
    "tool_calls": [
      {
        "type": "function",
        "id": "call_ea5621a2543f4a6fba92",
        "function": {
          "name": "bash",
          "arguments": {
            "command": "printf '%s\\n%s\\n' 'COMPLETE_TASK_AND_SUBMIT_FINAL_OUTPUT' '{\"prediction\": false}'"
          }
        }
      }
    ]
  }
}
\end{Verbatim}

\subsection{Output Example: Inspecting the Target-File Patch}
\label{app:teacher-output-example}

The following is the actual teacher proposal paired with the preceding input; all six fields are retained from the recorded proposal. The teacher requests the relevant patch section. Its explanation states only that absence from the base-version file does not establish absence from the reviewed version, without assuming the result of the proposed command.

\begin{Verbatim}[fontsize=\scriptsize,breaklines=true,breakanywhere=true]
{
  "kind": "bash",
  "command": "grep -n -A120 -B5 '^diff --git a/frontend/test-utils\\.tsx' /testbed/task/review.patch",
  "prediction": null,
  "explanation": "Inspect the focused patch section for the reviewed file. The repository is at the base commit and the patch is not applied, so the import's absence from the base file does not establish whether it is unused in the reviewed version.",
  "evidence": [
    {
      "message_index": 3,
      "quote": "\"repository_state\": \"base_commit; patch is not applied\""
    },
    {
      "message_index": 15,
      "quote": "returncode: 1\n<output>\n\n</output>"
    }
  ],
  "review_note": "The prior conclusion incorrectly treated base-version absence as patched-version evidence. The focused patch hunk should establish whether the added router namespace is used elsewhere in the reviewed file."
}
\end{Verbatim}

Here, \texttt{message\_index=3} refers to the tool observation containing the task JSON and its declaration that the repository is at the base commit, while \texttt{message\_index=15} refers to the search result with no match. The \texttt{prediction=null} value indicates that this action requests additional verification rather than submitting a final judgment.

\subsection{Teacher Configuration}
\label{app:teacher-configuration}

The teacher settings are:

\begin{Verbatim}[fontsize=\scriptsize]
teacher_model: gpt-5.6-sol
teacher_reasoning_effort: high
teacher_max_output_tokens: 4096
teacher_timeout: 180
teacher_followup_steps: 3
\end{Verbatim}

Table~\ref{tab:teacher-configuration} summarizes how these settings are used by the current implementation.

\begin{table}[H]
\centering
\footnotesize
\caption{Teacher API and orchestration configuration.}
\label{tab:teacher-configuration}
\renewcommand{\arraystretch}{1.08}
\begin{tabular}{p{0.28\linewidth}p{0.66\linewidth}}
\toprule
\textbf{Setting} & \textbf{Current implementation} \\
\midrule
Model & \texttt{gpt-5.6-sol}. \\
Reasoning effort & \texttt{high}, passed through \texttt{reasoning.effort}. \\
Maximum output tokens & \texttt{4096}, passed explicitly as \texttt{max\_output\_tokens}. The budget includes reasoning tokens and visible output. \\
Request timeout & 180 seconds, passed to the OpenAI SDK client; this is not a total budget for the full task. \\
SDK retries & \texttt{max\_retries=0}; the SDK does not retry requests automatically. \\
API call & \texttt{client.responses.create(...)}. \\
Output format & \texttt{text.format.type=json\_schema}, \texttt{name=teaching\_action}, and \texttt{strict=true}. \\
API storage flag & \texttt{store=false}; requests and responses are nevertheless recorded locally by the project. \\
Temperature / top-p & Neither value is set explicitly for this call; the teacher must not be described as using temperature 0. \\
Follow-up budget & For a given student prefix, the teacher first proposes one action and may then generate at most three additional actions along the same correction branch. Thus, each correction branch contains at most four teacher actions; these actions are generated sequentially rather than in a single response. \\
\bottomrule
\end{tabular}
\end{table}

\subsection{Reviewer Verification}
\label{app:reviewer-verification}

After a proposed teacher action is executed, an independent reviewer audits whether it is suitable as an action-level training target. Unlike the teacher, the reviewer observes the action's actual execution result and decides whether the resulting example should be retained.

\subsubsection{Reviewer Prompt}

The reviewer uses the following \texttt{REVIEW\_INSTRUCTION} verbatim:

\begin{Verbatim}[fontsize=\scriptsize,breaklines=true,breakanywhere=true]
Independently audit one proposed teaching action. You receive
the student's pre-action history, the action, its actual execution result (if a
command), and the PRIVATE human label. Do not approve merely because the label
matches. Check that the action advances technical/file verification, uses the
right code version, does not leak the answer key, and makes no factual claim that
required a future observation. References must actually support the claim, not
just occur somewhere in output. A no-match search can be a valid action. A final
judgment needs evidence, not an echoed label. Return needs_review if uncertain;
reject fabricated evidence, unsupported final judgments or invalid operations.
The action should be useful, but need not match the exact command you would use.
\end{Verbatim}

At request time, the implementation appends the same output-format instructions and complete \texttt{<student\_protocol>} used for the teacher, so the reviewer receives the student's judgment criteria and tool protocol.

\subsubsection{Input and Output Formats}

Table~\ref{tab:reviewer-input-format} summarizes the four reviewer-input fields.

\begin{table}[H]
\centering
\footnotesize
\caption{Fields supplied to the reviewer.}
\label{tab:reviewer-input-format}
\renewcommand{\arraystretch}{1.08}
\begin{tabular}{p{0.29\linewidth}p{0.65\linewidth}}
\toprule
\textbf{Field} & \textbf{Content} \\
\midrule
\path{teacher_private_label} & Human label for the current training task. \\
\texttt{student\_history} & Student-visible history before the teacher action. \\
\texttt{action} & Normalized teacher action, represented as an assistant message with one bash tool call. \\
\texttt{execution} & Actual \texttt{returncode} and output produced by executing the action. \\
\bottomrule
\end{tabular}
\end{table}

The reviewer returns strict JSON containing only \texttt{decision} and \texttt{reason}:

\begin{Verbatim}[fontsize=\scriptsize]
{
  "decision": "approve",
  "reason": "Explanation of the review decision."
}
\end{Verbatim}

\begin{table}[H]
\centering
\footnotesize
\caption{Reviewer decisions and their effects on the training pipeline.}
\label{tab:reviewer-decisions}
\renewcommand{\arraystretch}{1.08}
\begin{tabular}{p{0.16\linewidth}p{0.30\linewidth}p{0.44\linewidth}}
\toprule
\textbf{Decision} & \textbf{Meaning} & \textbf{Effect} \\
\midrule
\texttt{approve} & The action is suitable as a teaching example. & The record is marked approved and becomes eligible for inclusion in the training set. \\
\texttt{reject} & The action is invalid, unsupported by its evidence, or otherwise unsuitable. & The action is not exported, and the current correction branch terminates. \\
\bottomrule
\end{tabular}
\end{table}

For the ``Unused import'' example, the reviewer approves the patch-inspection action:

\begin{Verbatim}[fontsize=\scriptsize,breaklines=true,breakanywhere=true]
{
  "decision": "approve",
  "reason": "The action correctly distinguishes base from patched evidence and inspects the relevant patch section. Its output shows the reviewed import and the rest of the file's changes, allowing verification that `router` is not referenced. The command is valid, focused, useful, and makes no unsupported final judgment."
}
\end{Verbatim}

Reviewer requests reuse the teacher's model and request-level configuration in Appendix~\ref{app:teacher-configuration}; no separate reviewer model configuration is specified.

\clearpage
\section{Action-Level Distillation and Training}
\label{app:action-distillation}

\subsection{Training Record Format}
\label{app:action-training-format}

Each action-level supervised fine-tuning record contains three top-level fields: \texttt{prompt}, \texttt{completion}, and \texttt{tools}. The \texttt{prompt} contains the student instructions followed by its previously generated actions and actual tool observations. The \texttt{completion} contains exactly one reviewer-approved teacher action, represented as an assistant message with one bash tool call and an optional concise explanation. The \texttt{tools} field supplies the bash function schema used by both the prompt history and the target action.

The following abbreviated record uses the \textit{Unused import} example. Long task, patch, and file observations that already appear in Appendix~\ref{app:teacher-input-example} are replaced by explicit paper-only omission notices; the exported training record retains the full messages.

\begin{Verbatim}[fontsize=\scriptsize,breaklines=true,breakanywhere=true]
{
  "prompt": [
    {
      "role": "system",
      "content": "[Full student system prompt from Appendix A.1.1.]"
    },
    {
      "role": "user",
      "content": "Inspect /testbed/task/task.json and evaluate its target comment."
    },
    {
      "role": "assistant",
      "content": "I need to first examine the task file to understand what I'm evaluating.",
      "tool_calls": [
        {
          "type": "function",
          "id": "call_d2dc7d3ee7282ed7923e",
          "function": {
            "name": "bash",
            "arguments": {
              "command": "cat /testbed/task/task.json"
            }
          }
        }
      ]
    },
    {
      "role": "tool",
      "content": "[message_index=3]\nreturncode: 0\n<output>\n[Full task JSON observation omitted for presentation.]\n</output>",
      "tool_call_id": "call_d2dc7d3ee7282ed7923e"
    },
    {
      "role": "assistant",
      "content": "Now I need to check the actual patch file to see what changes were made and verify if the comment about unused import is correct.",
      "tool_calls": [
        {
          "type": "function",
          "id": "call_05af99c1e55acfbfe63b",
          "function": {
            "name": "bash",
            "arguments": {
              "command": "cat /testbed/task/review.patch"
            }
          }
        }
      ]
    },
    {
      "role": "tool",
      "content": "[message_index=5]\nreturncode: 0\n<output>\n[Long patch observation truncated by the tool and omitted here.]\n</output>",
      "tool_call_id": "call_05af99c1e55acfbfe63b"
    },
    {
      "role": "assistant",
      "content": "[Messages 6--14 contain the student's file search, base-file inspection, git checks, and final focused search.]"
    },
    {
      "role": "tool",
      "content": "[message_index=15]\nreturncode: 1\n<output>\n\n</output>",
      "tool_call_id": "call_01e89793b0d3d9b021a7"
    }
  ],
  "completion": [
    {
      "role": "assistant",
      "content": "Inspect the focused patch section for the reviewed file. The repository is at the base commit and the patch is not applied, so the import's absence from the base file does not establish whether it is unused in the reviewed version.",
      "tool_calls": [
        {
          "type": "function",
          "id": "call_0ecd1a0501cbde6a4fdf",
          "function": {
            "name": "bash",
            "arguments": {
              "command": "grep -n -A120 -B5 '^diff --git a/frontend/test-utils\\.tsx' /testbed/task/review.patch"
            }
          }
        }
      ]
    }
  ],
  "tools": [
    {
      "type": "function",
      "function": {
        "name": "bash",
        "description": "Execute a shell command in the isolated task container.",
        "parameters": {
          "type": "object",
          "properties": {
            "command": {
              "type": "string"
            }
          },
          "required": ["command"],
          "additionalProperties": false
        }
      }
    }
  ]
}
\end{Verbatim}

The prompt is used only as conditioning context. Completion-only loss is applied to the single teacher action in \texttt{completion}; student-history tokens and subsequent tool observations do not contribute to the objective.

\subsection{Training-Data Scale}
\label{app:action-training-scale}

Table~\ref{tab:action-training-scale} reports the action-level datasets used in the two iterative training rounds. Exported records are deduplicated by the tuple \texttt{(task\_id, history, action)}.

\begin{table}[H]
\centering
\small
\caption{Action-level training data used in the two rounds.}
\label{tab:action-training-scale}
\renewcommand{\arraystretch}{1.08}
\begin{tabular}{lrr}
\toprule
\textbf{Item} & \textbf{Round 1} & \textbf{Round 2} \\
\midrule
Action training examples & 4,622 & 4,646 \\
Training tasks represented & 701 & 714 \\
\bottomrule
\end{tabular}
\end{table}

\subsection{Optimization and Hardware Configuration}
\label{app:action-training-configuration}

We train only the LoRA adapter parameters and keep the base-model weights frozen. We tune the LoRA rank and alpha on the \textit{CRJudgeBench} validation split, selecting 16 and 32, respectively, and keep these values fixed for final evaluation on the test split. Table~\ref{tab:action-training-configuration} lists the training configuration used in both rounds.

\begin{table}[H]
\centering
\footnotesize
\caption{Action-level distillation training configuration.}
\label{tab:action-training-configuration}
\renewcommand{\arraystretch}{1.08}
\begin{tabular}{p{0.30\linewidth}p{0.64\linewidth}}
\toprule
\textbf{Setting} & \textbf{Value} \\
\midrule
Base model & \texttt{Qwen/Qwen3-Coder-30B-A3B-Instruct} \\
Updated parameters & LoRA parameters only; base-model weights are frozen. \\
LoRA target modules & \texttt{q\_proj}, \texttt{k\_proj}, \texttt{v\_proj}, and \texttt{o\_proj} \\
LoRA rank / alpha & 16 / 32 \\
LoRA dropout & 0.05 \\
Trainable parameters & 13,369,344 \\
Hardware & $4 \times$ NVIDIA H100 80GB GPUs \\
Distributed training & Distributed data parallel (DDP) \\
PyTorch & \texttt{2.14.0} \\
Transformers & \texttt{4.57.6} \\
PEFT & \texttt{0.18.0} \\
Accelerate & \texttt{1.12.0} \\
Learning rate & $5 \times 10^{-5}$ \\
Warmup ratio & 0.05 \\
Weight decay & 0.01 \\
Epochs per round & 3 \\
Checkpointing & Save every 50 optimizer steps and retain the three most recent checkpoints. \\
Final artifact & Save the final LoRA adapter separately in \path{adapter/}. \\
\bottomrule
\end{tabular}
\end{table}

\end{document}